\documentclass[12pt]{iopart}

\begin{document}

\comment[Comment on `Quantum inversion of cold atoms in a
microcavity']{Comment on `Quantum inversion of cold atoms in a
microcavity~: spatial dependence'}

\author{T Bastin and J Martin}
\address{Institut de Physique
Nucl\'eaire, Atomique et de Spectroscopie, Universit\'e de Li\`ege
au Sart Tilman, B\^at.\ B15, B - 4000 Li\`ege, Belgium}

\eads{\mailto{T.Bastin@ulg.ac.be}, \mailto{John.Martin@ulg.ac.be}}

\pacs{42.50.Pq, 42.50.Vk, 42.50.Ct, 32.80.Lg, 32.80.Pj}

\begin{abstract}
    In a recent work, Abdel-Aty and Obada [2002 \jpb {\bf 35} 807--13] analyzed the
    quantum inversion of cold atoms in a
    microcavity, the motion of the atoms being described quantum
    mechanically. Two-level atoms were assumed to interact with a single
    mode of the cavity, and the off-resonance case was considered
    (namely the atomic transition frequency is detuned from the
    single mode cavity frequency). We demonstrate in this paper
    that this case is incorrectly treated by these authors and we question
    therefore their conclusions.
\end{abstract}

\maketitle

\nosections

The interaction of cold atoms with microwave high-$Q$ cavities
(cold atom micromaser, also denominated mazer) has known these
last years an increasing interest since it has been demonstrated
by Scully \etal \cite{Scu96} that this interaction leads to a new
induced emission mechanism inside the cavity. In a recent work,
Abdel-Aty and Obada~\cite{Abd02a} extended the study of the mazer
by analyzing the quantum inversion of cold atoms interacting with
such cavities. In their model, two-level atoms are assumed to move
along the $z$-direction in the way to a cavity of length $L$. The
atomic center-of-mass motion is described quantum mechanically and
the atoms are coupled to a single mode of the quantized field
present in the cavity. The usual rotating-wave approximation is
made and contrary to previous works the general off-resonant case
is considered (the atomic transition and the cavity mode
frequencies are detuned). Their Hamiltonian reads
\begin{equation}
    \label{Hamiltonian}
        H = \frac{p_z^2}{2M} + \frac{\hbar \Delta}{2} \sigma_z + \hbar \omega \left( a^{\dagger} a + \frac{1}{2} \sigma_z \right) + \hbar \lambda  f(z) (a^{\dagger} \sigma + a
        \sigma^{\dagger}),
\end{equation}
where $p_z$ is the atomic center-of-mass momentum along the $z$
axis, $M$ the atomic mass, $\omega$ the cavity field mode
frequency, $\Delta$ is the detuning $\omega_0 - \omega$, with
$\omega_0$ the atomic transition frequency, $\sigma = |g \rangle
\langle e|$ ($|e\rangle$ and $|g\rangle$ are respectively the
upper and lower levels of the two-level atom), $\sigma_z$ is the
atomic inversion operator $|e \rangle \langle e| - |g \rangle
\langle g|$, $a$ and $a^{\dagger}$ are respectively the
annihilation and creation operators of the cavity radiation field,
$\lambda$ is the atom-field coupling strength and $f(z)$ is the
cavity field mode function.

In their studies, Abdel-Aty and Obada~\cite{Abd02a} consider an
atom initially in the excited state $|e\rangle$, and the field in
the arbitrary state $\sum_n D_n |n\rangle$, where $|n\rangle$ is
the photon number state. The center-of-mass motion is described by
the initial wave packet $\int dk \, G(k) e^{ikz} \theta(-z)$,
where $\theta(z)$ is the Heaviside step function, indicating that
the atoms are coming from the left part to the cavity. For a mesa
field mode ($f(z) = 1$, for $0 < z < L$), they
 find that the wave function,
$|\Psi(t)\rangle$, of the atom-field system is (for any time $t$)
\begin{eqnarray}
\label{psizt} \fl |\Psi(t)\rangle = \frac{1}{\sqrt{2}} \int dk \,
G(k) \exp\left( -i \frac{\hbar k^2 t}{2 M} \right) \sum_n D_n
\nonumber \\
 \times \left[ \left( \left[ e^{i k z} + (A^+_n + A^-_n)e^{-i k z} \right]\theta(-z) +  (B^+_n + B^-_n)e^{i k (z - L)} \theta(z - L) \right. \right. \nonumber \\
 + \left. \left\{ \alpha^+_n e^{i k^+_n z} + \beta^+_n
e^{-i k^+_n z} + \alpha^-_n e^{i k^-_n z} + \beta^-_n e^{-i k^-_n
z}\right\} \left[ \theta(z) - \theta(z - L) \right] \right) |e,
n\rangle \nonumber \\
 + \left( \left\{ \alpha^+_n e^{i k^+_n z} + \beta^+_n
e^{-i k^+_n z} - \alpha^-_n e^{i k^-_n z} - \beta^-_n e^{-i k^-_n
z} \right\} \left[ \theta(z) - \theta(z - L) \right] \right. \nonumber \\
 + \left. \left. (A^+_n - A^-_n)e^{-i k z}\theta(-z) +
(B^+_n - B^-_n)e^{i k (z - L)} \theta(z - L) \right) |g, n +
1\rangle \right],
\end{eqnarray}
with the coefficients $A^{\pm}_n$, $B^{\pm}_n$,
${\alpha}^{\pm}_n$, ${\beta}^{\pm}_n$ given by
\begin{eqnarray}
    A^{\pm}_n & = & i \Upsilon^{\pm}_n \sin(k^{\pm}_n L) \sin(\theta_n)
    \left[ \cos(k^{\pm}_n L) - i \delta^{\pm}_n \sin(k^{\pm}_n L)\right]^{-1}, \\
    B^{\pm}_n & = & \sin(\theta_n) e^{-i k L}
    \left[ \cos(k^{\pm}_n L) - i \delta^{\pm}_n \sin(k^{\pm}_n L)\right]^{-1}, \\
    {\alpha}^{\pm}_n & = & \frac{1}{2} \left( 1 + \frac{k}{k^{\pm}_n} \right) e^{-i k^{\pm}_n L} \sin(\theta_n)
    e^{-i k L}
    \left[ \cos(k^{\pm}_n L) - i \delta^{\pm}_n \sin(k^{\pm}_n L)\right]^{-1}, \\
    {\beta}^{\pm}_n & = & \frac{1}{2} \left( 1 - \frac{k}{k^{\pm}_n} \right) e^{i k^{\pm}_n L} \sin(\theta_n)
    e^{-i k L}
    \left[ \cos(k^{\pm}_n L) - i \delta^{\pm}_n \sin(k^{\pm}_n
    L)\right]^{-1},
\end{eqnarray}
and
\begin{eqnarray}
    k^{\pm}_n & = & \sqrt{k^2 \mp \gamma^2 \sqrt{\frac{\Delta^2}{4 \lambda^2}+(n +
    1)}} \,\, , \\
    \Upsilon^{\pm}_n & = & \frac{1}{2}\left( \frac{k^{\pm}_n}{k} -
    \frac{k}{k^{\pm}_n}\right), \\
    \delta^{\pm}_n & = & \frac{1}{2}\left( \frac{k^{\pm}_n}{k} +
    \frac{k}{k^{\pm}_n}\right), \\
    \theta_n & = & \tan^{-1} \left( \frac{\lambda \sqrt{n + 1}}{\sqrt{\frac{\Delta^2}{4} + \lambda^2 (n + 1)} - \frac{\Delta}{2}}
    \right),\label{thetan}
\end{eqnarray}
where $\gamma$ is defined by $(\hbar \gamma)^2/2 M = \hbar
\lambda$.

As clearly mentioned in Ref.~\cite{Abd02b}, if we denote
$|\Phi^{\pm}_n\rangle$ the atom-field dressed states
\begin{eqnarray}
    |\Phi^+_n\rangle & = & \cos \theta_n |g, n + 1\rangle + \sin
    \theta_n |e, n\rangle, \\
    |\Phi^-_n\rangle & = & -\sin \theta_n |g, n + 1\rangle + \cos
    \theta_n |e, n\rangle,
\end{eqnarray}
Abdel-Aty and Obada obtain Eqs.~(\ref{psizt})-(\ref{thetan}) of
the present comment by considering that the wave function
components $\Psi^{\pm}_n(z,t) = \langle z, \Phi^{\pm}_n |
\Psi(t)\rangle$ satisfy the Schr\"odinger equation,
\begin{equation}
\label{SchrodEq}
 i \hbar \frac{\partial}{\partial t} \Psi^{\pm}_n(z,t) = \left( \frac{-\hbar^2}{2 M} \frac{\partial^2}{\partial z^2} + V^{\pm}_n(z)
 \right) \Psi^{\pm}_n(z,t),
\end{equation}
with
\begin{equation}
V^{\pm}_n(z) = \pm \hbar \sqrt{\frac{\Delta^2}{4} + \lambda^2
f^2(z)(n + 1)} \,\, .
\end{equation}

Except at resonance ($\Delta = 0$), this assertion is wrong.
Contrary to what these authors claim, the wave function components
$\Psi^{\pm}_n(z,t)$ do not satisfy Eq.~(\ref{SchrodEq}) when a
detuning $\Delta$ is present. Their conclusions are therefore
questioned. Indeed, using the completeness relation
\begin{equation}
    1 = \int dz \sum_n \left( |z, \Phi^+_n\rangle \langle z, \Phi^+_n| +  |z, \Phi^-_n\rangle \langle z,
    \Phi^-_n|\right),
\end{equation}
and projecting the Schr\"odinger equation,
\begin{equation}
i \hbar \frac{d}{dt} |\Psi(t)\rangle = H |\Psi(t)\rangle,
\end{equation}
onto the dressed state basis $\left\{|z,
\Phi^{\pm}_n\rangle\right\}$ yields
\begin{equation}
    \fl
    i \hbar \frac{\partial}{\partial t} \Psi^{\pm}_n(z,t) = \int dz' \sum_{n'} \left( \langle z,
    \Phi^{\pm}_n | H | z', \Phi^+_{n'}\rangle \Psi^+_{n'}(z',t) + \langle z,
    \Phi^{\pm}_n | H | z', \Phi^-_{n'}\rangle \Psi^-_{n'}(z',t)
    \right),
\end{equation}
that is (after a straightforward calculation of the matrix
elements $\langle z, \Phi^{\pm}_n | H | z',
\Phi^{\pm}_{n'}\rangle$)
\begin{eqnarray}
\fl i\hbar\frac{\partial}{\partial t}\Psi^+_n(z,t)=\left[
  -\frac{\hbar^2}{2M}\frac{\partial^2}{\partial
    z^2} + \left(n+\frac{1}{2}\right)\hbar\omega-\cos 2\theta_n \frac{\hbar\Delta}{2}  + \hbar
    \lambda f(z) \sqrt{n+1}\sin2\theta_n\right]\Psi^+_n(z,t) \nonumber \\
+\left[\hbar \lambda f(z)
\sqrt{n+1}\cos2\theta_n+\sin2\theta_n\frac{\hbar\Delta}{2}\right]\Psi^-_n(z,t), \label{ssp}\\
\fl i\hbar\frac{\partial}{\partial t}\Psi^-_n(z,t)=\left[
  -\frac{\hbar^2}{2M}\frac{\partial^2}{\partial
    z^2}+\left(n+\frac{1}{2}\right)\hbar\omega+\cos 2\theta_n \frac{\hbar\Delta}{2}-\hbar
    \lambda
  f(z) \sqrt{n+1}\sin2\theta_n\right]\Psi^-_n(z,t) \nonumber \\
+\left[\hbar \lambda f(z)
\sqrt{n+1}\cos2\theta_n+\sin2\theta_n\frac{\hbar\Delta}{2}\right]\Psi^+_n(z,t).
\label{ssm}
\end{eqnarray}

We get, for each $n$, two coupled partial differential equations.
In presence of a detuning, the equations verified by the
components $\Psi^{\pm}_n(z,t)$ are much more complicated than at
resonance, and the atomic interaction with the cavity can no
longer be interpreted as an elementary scattering problem over two
potentials $V^+_n(z)$ and $V^-_n(z)$. Equations (\ref{ssp}) and
(\ref{ssm}) only reduce in the interaction picture to the simple
form (\ref{SchrodEq}) when $\Delta = 0$ ($\theta_n = \pi/4$).

We describe in details the effects of a detuning on the mazer
properties in Ref.~\cite{Bas02}. We show there that actually no
basis exists where Eqs.~(\ref{ssp}) and (\ref{ssm}) would separate
over the entire $z$ axis. Expressions for the probability of
finding the atoms in the excited state or in the ground state are
explicitly given for the mesa mode function. According to this
comment, our results differ significantly from those presented by
Abdel-Aty and Obada in Refs.~\cite{Abd02a} and \cite{Abd02b}.

We acknowledge the support of the Belgian Institut
Interuniversitaire des Sciences Nucl\'eaires (IISN).

\end{document}